\documentclass[10pt]{article}

\usepackage[margin=1in]{geometry}
\usepackage[toc,page]{appendix}
\usepackage{graphicx} % Required for inserting images
\usepackage{epsfig}
\usepackage{amsfonts}
\usepackage{amssymb}
\usepackage{amstext}
\usepackage{amsmath}
\usepackage{xspace}
\usepackage{hyperref}
\usepackage[T1, OT1]{fontenc}
\DeclareTextSymbolDefault{\dh}{T1}
\usepackage{fullpage}
\usepackage{enumitem}                     
\usepackage{titlesec}
\usepackage{amsthm}
\usepackage{natbib}
\usepackage{mathabx}
\usepackage{bbm}
\usepackage{xcolor}
\usepackage{mdframed}
\usepackage{centernot}
\usepackage{lmodern}
\usepackage{mdframed}
\usepackage{multirow}
\usepackage{floatrow}
\usepackage{setspace}
\usepackage{float}
\usepackage{authblk}

\titleformat*{\section}{\bfseries}
\titleformat*{\subsection}{\bfseries}
\titleformat*{\subsubsection}{\bfseries}
\titleformat*{\paragraph}{\bfseries}
\titleformat*{\subparagraph}{\bfseries}

\setenumerate[0]{label=(\alph*)}

\newcommand{\X}{\ensuremath{\mathbf{X}}}
\newcommand{\x}{\ensuremath{\mathbf{x}}}
\newcommand{\homega}{\hat\omega}

\renewcommand{\P}[2]{\ensuremath{{\text{Pr}}_{#1}\left(#2\right)}}

\newcommand{\E}[2]{\ensuremath{{\mathbb E}_{#1}\left[#2\right]}}

\newcommand{\p}[1]{\ensuremath{\left(#1\right)}}

\newcommand{\sbr}[1]{\ensuremath{\left[#1\right]}}

\providecommand{\keywords}[1]
{
	\normalsize	
	\noindent \textbf{Keywords:} #1
}

\title{Addressing Population Heterogeneity \protect\\ for HIV Incidence Estimation Based on Recency Test}
\author[1]{Qi Wang}
\author[2,3]{Ann Duerr}
\author[2,3]{Fei Gao}
\affil[1]{Department of Statistics, University of Washington, Seattle, Washington, USA}
\affil[2]{Vaccine and Infectious Disease Division, Fred Hutchinson Cancer Research Center, Seattle, Washington, USA}
\affil[3]{Public Health Sciences Division, Fred Hutchinson Cancer Research Center, Seattle, Washington, USA}
\date{}
\begin{document}

\maketitle

\begin{abstract}
    Cross-sectional HIV incidence estimation leverages recency test results to determine the HIV incidence of a population of interest, where recency test uses biomarker profiles to infer whether an HIV-positive individual was ``recently'' infected. This approach possesses an obvious advantage over the conventional cohort follow-up method since it avoids longitudinal follow-up and repeated HIV testing. In this manuscript, we consider the extension of cross-sectional incidence estimation to estimate the incidence of a different target population addressing potential population heterogeneity.
    We propose a general framework that incorporates two settings: one with the target population that is a subset of the population with cross-sectional recency testing data, e.g., leveraging recency testing data from screening in active-arm trial design, and the other with an external target population.
    We also propose a method to incorporate HIV subtype, a special covariate that modifies the properties of recency test, into our framework. Through extensive simulation studies and a data application, we demonstrate the excellent performance of the proposed methods. We conclude with a discussion of sensitivity analysis and future work to improve our framework.
    
    \vspace{1em}
    
\end{abstract}
% \textbf{Keywords:} HIV, HIV subtype, incidence, population heterogeneity, recency test.

\clearpage

\section{Introduction}
Recent years have witnessed the broad utility of cross-sectional incidence estimation, i.e., estimating HIV incidence utilizing data from a cross-sectional sample.
Such an approach leverages recency test results, which are from an algorithm with single or multiple biomarkers to classify whether an HIV-positive individual was infected ``recently'', e.g., infected within the past year. 
Cross-sectional incidence estimators 
\citep{kaplan1999snapshot, kassanjee2012new} use a representative cross-sectional sample to estimate the incidence in a target population, under assumptions regarding the epidemic in the target population and properties of the recency test algorithm \citep{gao2022statistical}.
This approach does not require longitudinal follow-up and repeated HIV testing, which confers it an obvious advantage over the traditional cohort follow-up approach.

Cross-sectional incidence estimation has also been studied to serve as a comparator in evaluating the prevention efficacy of an experimental pre-exposure prophylaxis (PrEP) product in an HIV prevention trial \citep{gao2021sample}.
%Efficacy evaluation of investigative PrEP agents was traditionally done by placebo-controlled trials (e.g., \cite{grant2010preexposure}, \cite{thigpen2012antiretroviral}, \cite{marrazzo2013pre}, and \cite{van2012preexposure}), where individuals are randomly assigned to receive either the experimental product or placebo.
%After the success of Tenofovir/emtricitabine (TDF/FTC), current practice is to conduct an active-control trial \citep{landovitz2021cabotegravir,delany2022cabotegravir}, where individuals are randomized to receive either the experimental product or an active-control product that has been demonstrated efficacious in a previous trial.
The current practice for efficacy evaluation of investigative PrEP agents is through an active-control trial (e.g., \cite{landovitz2021cabotegravir} and \cite{delany2022cabotegravir}), where individuals are randomized to receive either the experimental product or an active-control product that has been demonstrated efficacious in a previous trial.
With the development of highly efficacious PrEP agents in the past years, e.g., cabotegravir, conducting active-control trials may not be feasible, due to the large number of participants and the long follow-up time required by fully powered active-control trials.
In addition, the absolute efficacy of the experimental PrEP product compared to placebo is not directly inferred in such trials.
One promising alternative is to supplement the active-arm trial with a ``counterfactual placebo'' incidence estimated by a cross-sectional incidence estimator applied to the screening population \citep{gao2021sample}. At the screening stage, individuals are tested for HIV status and (sometimes a proportion of) HIV-positive individuals are tested by recency test. A subset of eligible HIV-negative individuals who consent will be enrolled in the trial and followed up. The efficacy of the investigative product is assessed by comparing the estimated incidence in the active arm and the estimated counterfactual placebo incidence, under the assumption that screening and trial populations are similar. However, such an assumption is unlikely to hold in practice. For example, inclusion and exclusion criteria may lead to differential distributions of HIV risk factors between the screening and trial populations.
Directly applying the existing approach fails to account for such population heterogeneity and may lead to biased results.

Facilitating active-arm trial design with cross-sectional incidence estimation is in line with the framework of generalizability and transportability in the statistical literature, which aims to extend a causal estimand from randomized or observational data of a study population to a different target population \citep{degtiar2023review}. Generalizability focuses on the setting where the study population is a subset of the target population, while for transportability, the study population is external to the target population.
%Similarly, using cross-sectional incidence estimation to supplement active-arm trials is to extend the incidence estimate of the cross-sectional screening population to a target population that is internal. 
Many methods have been proposed in the field of generalizability and transportability, including matching \citep{stuart2011use,bennett2020building}, 
weighting-based methods \citep{flores2013comparing, lesko2017generalizing, westreich2017transportability, dahabreh2019generalizing, dahabreh2020extending}, % \citep{flores2013comparing, lesko2017generalizing, westreich2017transportability, correa2018generalized, dahabreh2019generalizing, dahabreh2020extending, westreich2017transportability, dahabreh2020extending}
outcome regressions \citep{flores2013comparing, dahabreh2019generalizing, dahabreh2020extending, kern2016assessing}, and combined propensity score and outcome regression methods \citep{dahabreh2019generalizing, dahabreh2020extending, rudolph2017robust, schmid2022comparing, dong2020using}. However, those approaches cannot be directly applied to addressing population heterogeneity with cross-sectional incidence estimation. Existing approaches usually utilize observed outcomes in the study population to infer some quantities related to the same (unobserved) outcomes in the target population, leveraging similarity between the populations. It is unclear how such ``outcomes'' should be defined in our setting with cross-sectional HIV incidence estimation. In addition, in our setting, the ``study population'' includes both HIV positive and negative individuals, while the ``target population'' includes only HIV negative individuals. It is not obvious how the ``similarity'' of the two populations can be established.

In this paper, we will develop a framework to address population heterogeneity for HIV incidence estimation based on recency test. We consider two settings: one with a target population that is a subset of the population with cross-sectional recency testing data, e.g., leveraging recency testing data from screening in active-arm trial design, and the other with an external target population. The latter setting includes the application of incidence estimation for a target population, making use of historical recency testing data collected in the same geographic area. Such background incidence estimate obtained can facilitate the design of a trial to be conducted in the target population.
%In addition, we propose estimators when the recency testing data are from a population with a mixture of subtypes, which is a special factor that contributes to population heterogeneity and affects the property of the recency test.
We also propose a method to incorporate HIV subtype, a special covariate that modifies the properties of recency test, into our framework.
The rest of the paper is organized as follows.
We describe our framework and proposed estimators in Section 2. In Section 3, we evaluate the performance of the proposed approach through extensive simulation studies. Section 4 describes an analysis of data from the Sabes study, a treatment-as-prevention intervention study in Lima, Peru. We conclude with a discussion in Section 5.

\section{Method}
\subsection{Cross-sectional incidence estimation}
\label{subsec: Cross-sectional incidence estimation by recency tests}
Two commonly used estimators for cross-sectional incidence estimation are the snapshot estimator by \cite{kaplan1999snapshot} and an adjusted estimator by \cite{kassanjee2012new}. Here we focus on the adjusted estimator by \cite{kassanjee2012new}, although the same methodology would apply to the snapshot estimator. Let $T^*$ denote a cutoff time, such that ``recent'' infection is defined as an infection occurring less than $T^*$ time units ago. The incidence estimator adopts two characteristics of a recency test based on this cutoff: mean duration of infection (MDRI) $\Omega_{T^*}$ and false-recent rate (FRR) $\beta_{T^*}$. MDRI captures the average duration of infection among those infected less than $T^*$ time units ago and classified as test-recent; FRR is the probability of recency-test-positive for a randomly selected individual infected more than $T^*$ time units ago. Let $N_{\text{pos}}$, $N_{\text{neg}}$, and $N_{\text{rec}}$ be the numbers of HIV-positive, HIV-negative individuals, and individuals tested recent, respectively, in the cross-sectional sample. The estimator for HIV incidence by \cite{kassanjee2012new} is given by
\begin{align}
\label{eq: original cross sectional incidence estimator}
    % \Tilde{\lambda}=\frac{N_{\text{rec}}-N_{\text{pos}}\hat{\beta}_{T^*}}{N_{\text{neg}}(\hat{\Omega}_{T^*}- \hat{\beta}_{T^*} T^*)},
    \Tilde{\lambda}=\frac{N_{\text{rec}}-N_{\text{pos}}\hat{\beta}_{T^*}}{N_{\text{neg}}(\hat{\Omega}_{T^*}- \hat{\beta}_{T^*} T^*)},
\end{align}
where $\hat{\Omega}_{T^*}$ and $\hat{\beta}_{T^*}$ are estimates for MDRI and FRR from external studies with individuals with known infection time (see \cite{gao2022statistical} for more details).

This cross-sectional incidence estimator can also be written in the form of individual contributions.
Write $N = N_{\text{pos}} + N_{\text{neg}}$ as the size of the cross-sectional sample.
For individual $i=1,\dots,N$, let $D_i$ be the indicator for HIV status and $R_i$ be the indicator of whether individual $i$ is test-positive for the recency test (test-recent). The incidence estimator $\Tilde{\lambda}$
% (\ref{eq: original cross sectional incidence estimator})
can also be written as
\begin{align}
   \Tilde{\lambda}=\frac{N^{-1}\sum_{i=1}^N D_i(R_i-\hat{\beta}_{T^*})}{N^{-1}\sum_{i=1}^N (1-D_i)(\hat{\Omega}_{T^*}- \hat{\beta}_{T^*} T^*)}.
   \label{eq: cross sectional incidence esimator}
\end{align}
\cite{gao2022statistical} has shown that this estimator is consistent if two key assumptions hold: the HIV incidence and prevalence in the population are constant over a time period prior to the cross-sectional sample; and the duration-specific test-recent probability function of the recency test is constant in the tail. We note that \cite{gao2022statistical} also implicitly assumes that the HIV incidence (and prevalence) is homogeneous among individuals and the estimator is consistent for such homogeneous incidence. In our paper, we consider heterogeneous prevalence and incidence among individuals. Specifically, let $\X$ denote a collection of covariates that modify HIV prevalence and incidence. \footnote{We assume that $\X$ captures all the factors that modify HIV prevalence and incidence. In other words, there is no unmeasured confounder.}
The first step of our method development is to derive what the estimator $\Tilde\lambda$ is consistent for when prevalence and incidence are heterogeneous.

Let $T$ be the calendar time of HIV infection for an individual. At any calendar time $t$, let the prevalence and incidence of an individual with covariates $\X=\x$ be $p(t|\x)=\P{}{T\le t|\X=\x}$ and $\lambda(t|\x)=\lim_{dt\rightarrow 0} \frac{1}{dt} \P{}{t\le T<t+dt|T\ge t,\X=\x}$, respectively.
%Let $P_{\text{rec}}(t|\x)$ denote the probability that an individual with $\X=\x$ is classified as test-recent at time $t$ by the recency test.
Similar to \cite{gao2022statistical}, we assume $p(t|\x)$ and $\lambda(t|\x)$ are constant over time.
Hereafter, we can simplify our notations by $p(t|\x):=p(\x)$ and $\lambda(t|\x):=\lambda(\x)$. Without loss of generality, we assume covariates $\X$ are continuous and let $g(\x)$ be the probability density function of $\X$ in the population where the cross-sectional sample is from. We also assume that $\X$ does not modify the properties of the recency test, while the incorporation of recency test modifiers, i.e., HIV subtypes, will be discussed in Section \ref{subsec: subtype}.

Let $\phi(u)$ be the test-recent probability of an individual with HIV infection duration of $u$. By the definition of MDRI, $\Omega_{T^*}=\int_0^{T^*}\phi(u)du$. One of the two key assumptions from \cite{gao2022statistical} is that $\phi(u)=\beta_{T^*}$ when $u\ge T^*$. Suppose the cross-sectional time of the recency testing data is time $t$. Then,
\begin{align*}
    &\P{}{D_i=1,R_i=1|\X_i=\x_i}\\
    =&\int_0^\infty \phi(u) \P{}{T=t-u|\X_i=\x_i} du\\
    =&\int_0^\infty \phi(u) \P{}{T=t-u|T\le t,\X_i=\x_i} \P{}{T\le t|\X_i=\x_i}du\\
    =&p(\x_i)\sbr{\int_0^{T^*} \phi(u) \P{}{T=t-u|T\le t,\X_i=\x_i}du+\int_{T^*}^{\infty} \beta_{T^*} \P{}{T=t-u|T\le t,\X_i=\x_i}du}\\
    =&p(\x_i)\sbr{\int_0^{T^*} \phi(u) \P{}{T=t-u|T\le t,\X_i=\x_i}du+\beta_{T^*} \p{1-\int_0^{T^*}\P{}{T=t-u|T\le t,\X_i=\x_i}du}}.
\end{align*}
Considering that
\begin{align*}
    \P{}{T=t-u|T\le t,\X_i=\x_i}=\frac{\P{}{T=t-u|T\ge t-u,\X_i=\x_i}\P{}{T\ge t-u|\X_i=\x_i}}{\P{}{T\le t|\X_i=\x_i}}=\frac{\lambda(\x_i)(1-p(\x_i))}{p(\x_i)},
\end{align*}
we have
\begin{align*}
    \P{}{D_i=1,R_i=1|\X_i=\x_i}=&p(\x_i)\sbr{\Omega_{T^*}\frac{\lambda(\x_i)(1-p(\x_i))}{p(\x_i)}+\beta_{T^*}\p{1-T^*\frac{\lambda(\x_i)(1-p(\x_i))}{p(\x_i)}}}\\
    =&(\Omega_{T^*}-\beta_{T^*}T^*)\lambda(\x_i)(1-p(\x_i))+\beta_{T^*}p(\x_i).
\end{align*}
Now for the numerator of \eqref{eq: cross sectional incidence esimator}, 
\begin{align*}
    \frac{1}{N}\sum_{i=1}^N D_i(R_i-\hat{\beta}_{T^*})\overset{\text{p}}{\rightarrow}&{\E{}{D_i(R_i-\hat{\beta}_{T^*})}}=\E{}{\E{}{D_i(R_i-\hat{\beta}_{T^*})|\X_i}}\\
    =&\E{}{\E{}{D_iR_i|\X_i} - \E{}{D_i\hat{\beta}_{T^*}|\X_i}}\\
    =&\E{}{(\Omega_{T^*}-\beta_{T^*}T^*)\lambda(\X_i)(1-p(\X_i))+\beta_{T^*}p(\X_i) - \beta_{T^*}p(\X_i)}\\
    =&(\Omega_{T^*}-\beta_{T^*}T^*)\E{}{\lambda(\X_i)(1-p(\X_i))},
\end{align*}
where $\overset{\text{p}}{\rightarrow}$ denotes convergence in probability. Similarly, for the denominator of \eqref{eq: cross sectional incidence esimator},
\begin{align*}
     \frac{1}{N}\sum_{i=1}^N (1-D_i)(\hat{\Omega}_{T^*}- \hat{\beta}_{T^*} T^*)\overset{\text{p}}{\rightarrow}&\E{}{(1-D_i)(\hat{\Omega}_{T^*}-\hat{\beta}_{T^*}T^*)}=\E{}{\E{}{(1-D_i)(\hat{\Omega}_{T^*}-\hat{\beta}_{T^*}T^*)|\X_i}}\\
    =&\E{}{(\Omega_{T^*}-\beta_{T^*}T^*)\E{}{(1-D_i)|\X_i}}=(\Omega_{T^*}-\beta_{T^*}T^*)\E{}{1-p(\X_i)}.
\end{align*}
Then, we get
\begin{align}
\label{eq:asymptotic cross sectional incidence estimator}
    \Tilde{\lambda}\overset{\text{p}}{\rightarrow} \frac{\E{}{\lambda(\X)(1-p(\X))}}{\E{}{1-p(\X)}}=\frac{\int \lambda(\x)(1-p(\x))g(\x)d\x}{\int (1-p(\x))g(\x)d\x}.
\end{align}
Equation \eqref{eq:asymptotic cross sectional incidence estimator} implies that the incidence estimator is consistent for the average incidence of HIV-negative individuals in the population with the cross-sectional recency testing data. This is in line with modern epidemiological studies on prospective HIV incidence where researchers focus on the HIV incidence estimate of the HIV-seronegative population at baseline \citep{kroidl2016effect, justman2017swaziland, kagaayi2019impact}.

\subsection{Extension of cross-sectional incidence estimate accounting for population heterogeneity}
\label{subsec: incidence two populations}

We consider extending the cross-sectional incidence estimate to a target population that naturally is only composed of HIV-negative individuals.
%As argued in the previous section, it is only sensible to consider HIV-negative populations for incidence estimation.
In the following, we will call the population with the cross-sectional recency testing data as ``cross-sectional population''.
%This requirement is realistic in the sense that our goal is to design a trial for the target population and only HIV-negative individuals can be enrolled into the trial, which indicates that our target population is exactly an HIV-negative population.
Let $h(\x)$ be the probability density function of covariates $\X$ in the target population.

% The incidence estimator for population $1$ can be written as
% \begin{align}
%     \Tilde{\lambda}_1=\frac{\sum_{i=1}^N I(S_i=1)D_i(R_i-\hat{\beta}_{T^*})}{\sum_{i=1}^N I(S_i=1)(1-D_i)(\hat{\Omega}_{T^*}- \hat{\beta}_{T^*} T^*)}
% \end{align}

We aim to build a consistent estimator for the average incidence of the target population, i.e., $\int \lambda(\x) h(\x)d\x$, by modifying \eqref{eq: cross sectional incidence esimator}. 
Motivated by the inverse probability weighting approach% \citep{flores2013comparing, lesko2017generalizing, westreich2017transportability, correa2018generalized, dahabreh2019generalizing, dahabreh2020extending, westreich2017transportability, dahabreh2020extending}
, we propose to re-weight the numerator and denominator of \eqref{eq: cross sectional incidence esimator}. Specifically, we propose the extended estimator given by
\begin{align}
\label{eq: random trial placebo incidence estimator}
    \hat{\lambda}_{\text{extend}}=&\frac{N^{-1}\sum_{i=1}^N \omega(\X_i) D_i(R_i-\hat{\beta}_{T^*})}{N^{-1}\sum_{i=1}^N \omega(\X_i)(1-D_i)(\hat{\Omega}_{T^*}- \hat{\beta}_{T^*} T^*)}.
\end{align}
For the numerator of \eqref{eq: random trial placebo incidence estimator},
\begin{align*}
    \frac{1}{N}\sum_{i=1}^N \omega(\X_i) D_i(R_i-\hat{\beta}_{T^*})\overset{\text{p}}{\rightarrow}&\E{}{\omega(\X_i)D_i(R_i-\hat{\beta}_{T^*})}=\E{}{\E{}{\omega(\X_i)D_i(R_i-\hat{\beta}_{T^*})|\X_i}}\\
    =&\E{}{\omega(\X_i)\p{\E{}{D_iR_i|\X_i} - \E{}{D_i\hat{\beta}_{T^*}|\X_i}}}\\
    % =\E{}{\omega(\X_i)\p{P_{\text{rec}}(\X_i) - \beta_{T^*}p(\X_i)}}\\
    % =&\E{}{\omega(\X_i)\sbr{(\Omega_{T^*}-\beta_{T^*}T^*)\lambda(\X_i)(1-p(\X_i))+\beta_{T^*}p(\X_i) - \beta_{T^*}p(\X_i)}}\\
    =&(\Omega_{T^*}-\beta_{T^*}T^*)\E{}{\omega(\X_i)\lambda(\X_i)(1-p(\X_i))}.
\end{align*}
Similarly, we can show that for the denominator of \eqref{eq: random trial placebo incidence estimator},
\begin{align*}
    \frac{1}{N}\sum_{i=1}^N \omega(\X_i)(1-D_i)(\hat{\Omega}_{T^*}- \hat{\beta}_{T^*} T^*)\overset{\text{p}}{\rightarrow}(\Omega_{T^*}-\beta_{T^*}T^*)\E{}{\omega(\X_i)\p{1-p(\X_i)}}.
\end{align*}
This implies that
\begin{align*}
    \hat{\lambda}_{\text{extend}} \overset{\text{p}}{\rightarrow} \frac{\E{}{\omega(\X_i)\lambda(\X_i)(1-p(\X_i))}}{\E{}{\omega(\X_i)\p{1-p(\X_i)}}}=\frac{\int \omega(\x)\lambda(\x)(1-p(\x))g(\x) d\x}{\int \omega(\x)(1-p(\x))g(\x) d\x}.
\end{align*}
Therefore, to make the extended estimator consistent for $\int \lambda(\x) h(\x)d\x$, the weight should take the form
\begin{align}
    \omega(\x)\propto &\frac{h(\x)}{(1-p(\x))g(\x)} \label{eq: weight}.
\end{align}
The weight is proportional to the ratio of the densities of covariates $\X$ in the target population and the HIV-negative individuals in the cross-sectional population.
Such a weight is intuitive, given that the original incidence estimator \eqref{eq: cross sectional incidence esimator} pertains to the average incidence for the HIV-negative individuals in the cross-sectional population, and we wish to extend such incidence to the target population.

Next, we will discuss how to implement estimator \eqref{eq: random trial placebo incidence estimator} by replacing $\omega(\X)$ with its estimator in two different scenarios.

\subsubsection{Weight estimation when the target population is external}
\label{subsubsec: weight two independent populations}
We first consider the case when the target population is external to the cross-sectional population. For example, to design an HIV prevention trial, researchers may utilize the cross-sectional data on recency testing collected in the same geographic area to infer a background incidence for the trial population and aid sample size calculation.

To estimate the weight, we define a pooled population by combining the cross-sectional population and the target population.
Let $S_i$ indicate which population individual $i$ is from, where $S_i = 1$ indicates the target population and $S_i=0$ indicates the cross-sectional population.
Then, the weight \eqref{eq: weight} can be written as
\begin{align*}
    \omega(\x)\propto \frac{h(\x)}{\p{1-p(\x)}g(\x)}=&\frac{f(\x|S_i=1)}{\P{}{D_i=0|S_i=0,\X_i=\x}f(\x|S_i=0)}\\
    =&\frac{\P{}{S_i=1|\X_i=\x}f(\x)/\P{}{S_i=1}}{\P{}{D_i=0|S_i=0,\X_i=\x}\P{}{S_i=0|\X_i=\x}f(\x)/\P{}{S_i=0}}\\
    =&\frac{\P{}{S_i=1|\X_i=\x}}{\P{}{D_i=0|S_i=0,\X_i=\x}\P{}{S_i=0|\X_i=\x}}\frac{\P{}{S_i=0}}{\P{}{S_i=1}}\\
    \propto&\frac{\P{}{S_i=1|\X_i=\x}}{\P{}{S_i=0,D_i=0|\X_i=\x}},
\end{align*}
where we let $f(\x|\cdot)$ denote the generic notation for the probability density function of covariates $\X$ in a population.
That is, we can apply the weight
\begin{align*}
% \label{eq: weight when target population is external}
    \omega_{\text{ext}}(\x)=\frac{\P{}{S_i=1|\X_i=\x}}{\P{}{S_i=0,D_i=0|\X_i=\x}},
\end{align*}
which can be estimated by any statistical learning model, for example, logistic regression, using the pooled dataset including all individuals in the target population and HIV-negative individuals in the cross-sectional population.
The incidence in the target population can then be estimated by
\begin{align}
\label{eq: incidence_external_target}
    \hat{\lambda}_{\text{ext}}=&\frac{N^{-1}\sum_{i=1}^N \homega_{\text{ext}}(\X_i) D_i(R_i-\hat{\beta}_{T^*})}{N^{-1}\sum_{i=1}^N \homega_{\text{ext}}(\X_i)(1-D_i)(\hat{\Omega}_{T^*}- \hat{\beta}_{T^*} T^*)},
\end{align}
where $\homega_{\text{ext}}(\x)$ is the estimated weight.
Note that even though the weight is constructed based on HIV-negative individuals only, it is applied to both HIV-negative and HIV-positive individuals in the cross-sectional population.

% It should be emphasized that at the stage of designing a prevention trial, though we don't have the data of $S_i$ and $\X_i$ since we have not recruited any individuals yet, we may assume the distribution of covariates in the trial population is similar to an existing trial. Then, we can estimate the weight and further estimate the background incidence of the trial population based on the cross-sectional recency testing data and the auxiliary trial data.

\subsubsection{Weight estimation when the target population is internal}
\label{subsubsec: weight active-arm trial design}
% Evaluation of the efficacy of investigative PrEP agents is typically done by randomized active-control trials , where individuals are randomly assigned to either treatment arm with PrEP products or control arm. However, in some cases, it is not feasible to conduct such trials due to the large number of participants and the long follow-up time required by fully powered active-control trials. One promising alternative is to use the counterfactual placebo incidence estimated by the cross-sectional estimator and to conduct an active-arm trial \citep{gao2021sample}. At the screening stage, the individuals are tested for HIV status and (sometimes a proportion of) HIV-positive individuals are tested by recency assay. A subset of HIV-negative individuals who consent to be enrolled will be enrolled in the trial and followed up as a cohort. The efficacy of the investigative product is assessed by comparing the estimated incidence in the active arm and the counterfactual placebo incidence. \textcolor{red}{However, to account for the heterogeneity between the populations at the screening stage and the trial stage induced by inclusion and exclusion criteria and individuals' consent, we also need to modify the cross-sectional incidence estimator by plugging in a weight.}

Now we consider the case when the target population is a subset of the cross-sectional population, for example, when the cross-sectional population is the screening population in an active-arm trial, and we wish to infer a counterfactual placebo incidence for the trial population accounting for potential heterogeneity.
Here, $S_i$ is defined as the indicator of whether an individual $i$ in the cross-sectional sample also belongs to the target population.
Based on the definition of $S_i$, it is evident that $h(\x)=g(\x|D_i=0,S_i=1)$.
The weight \eqref{eq: weight} can be written as
\begin{align*}
    \omega(\x)\propto \frac{h(\x)}{\p{1-p(\x)}g(\x)}=&\frac{g(\x|D_i=0,S_i=1)}{\P{}{D_i=0|\X_i=\x}g(\x)}\\
    =&\frac{\P{}{D_i=0,S_i=1|\X_i=\x}g(\x)/\P{}{D_i=0,S_i=1}}{\P{}{D_i=0|\X_i=\x}g(\x)}\\
    \propto&\P{}{S_i=1|D_i=0,\X_i=\x}.
\end{align*}
Similar to Section \ref{subsubsec: weight two independent populations}, we may apply the weight
\begin{align*}
    \omega_{\text{int}}(\x)=\P{}{S_i=1|D_i=0, \X_i=\x},
\end{align*}
which is the probability that an HIV-negative individual with covariates $\x$ is included in the target population.
The incidence in the target population can then be estimated by
\begin{align}
\label{eq: incidence_internal_target}
    \hat{\lambda}_{\text{int}}=&\frac{N^{-1}\sum_{i=1}^N \homega_{\text{int}}(\X_i) D_i(R_i-\hat{\beta}_{T^*})}{N^{-1}\sum_{i=1}^N \homega_{\text{int}}(\X_i)(1-D_i)(\hat{\Omega}_{T^*}- \hat{\beta}_{T^*} T^*)},
\end{align}
where $\homega_{\text{int}}(\x)$ is an estimate for the weight $\omega_{\text{int}}(\x)$.

\subsection{incorporating HIV subtypes}
\label{subsec: subtype}
HIV has multiple distinct subtypes: for example, subtype B is the most prevalent viral strain in North America \citep{oster2017increasing}, Europe \citep{sallam2017molecular,tumiotto2018diversity,volz2018molecular,hebberecht2018frequency,alexiev2018origin} and Australia \citep{castley2017national}, whereas subtype C predominates in Southern Africa \citep{ngcapu2017characterization, sivay2018hiv}.
HIV subtype can be viewed as a special covariate since it modifies the properties of recency tests, such that the MDRI and FRR for the same recency test are different for individuals with different HIV subtypes.
Current and future HIV prevention trials are commonly worldwide. For example, HVTN 704/HPTN 085 enrolls men and transgender people who have sex with men in North America, South America, and Switzerland \citep{corey2021two}. 
It is then crucial to propose a cross-sectional incidence estimator based on recency test results of mixed subtypes.
In this section, we consider the scenario where the cross-sectional sample involves $K$ HIV subtypes and modify the estimators \eqref{eq: original cross sectional incidence estimator}, \eqref{eq: incidence_external_target}, and \eqref{eq: incidence_internal_target} to account for HIV subtypes.

\subsubsection{Modifying cross-sectional incidence estimator incorporating HIV subtypes}
\label{subsubsec: cross-sectional incidence estimator subtype}

To incorporate HIV subtypes in cross-sectional incidence estimation, existing literature proposed to use the weighted MDRI and FRR based on the sample proportions of each subtype \citep{parkin2023facilitating, sempa2019performance}.
Such an incidence estimator is given by
\begin{align}
\label{eq: incid estimator 1 subtype}
    \Tilde{\lambda}^{\text{sub}}_{\text{old}}=\frac{N_{\text{rec}}-N_{\text{pos}}\sum_{j=1}^K \hat{\pi}_j\hat{\beta}_{T^*,j}}{N_{\text{neg}}\p{\sum_{j=1}^K \hat{\pi}_j\hat{\Omega}_{T^*,j}- \sum_{j=1}^K \hat{\pi}_j\hat{\beta}_{T^*,j} T^*}},
\end{align}
where $\hat{\Omega}_{T^*,j}$ and $\hat{\beta}_{T^*,j}$ are the estimated MDRI and FRR of HIV subtype $j$, respectively, $\hat{\pi}_j = N_j/N$ is the estimated proportion for subtype $j$, and $N_j$ is the number of individuals in the cross-sectional sample corresponding to subtype $j$. As we have shown in Appendix \ref{app: proofs of cross-sectional subtype},
\begin{align*}
    \Tilde{\lambda}^{\text{sub}}_{\text{old}} \overset{\text{p}}{\rightarrow}\,& \frac{\sum_{j=1}^K \sbr{(\Omega_{T^*,j}-\beta_{T^*,j}T^*)(1-p(j))\pi_j\lambda(j)+p(j)\pi_j\beta_{T^*,j}} - \p{\sum_{j=1}^K p(j)\pi_j}\p{\sum_{j=1}^K \pi_j\beta_{T^*,j}}}{\p{1-\sum_{j=1}^K p(j)\pi_j}\p{\sum_{j=1}^K \pi_j(\Omega_{T^*,j}-\beta_{T^*,j}T^*)}},
\end{align*}
where $\pi_j$, $p(j)$, $\lambda(j)$, $\Omega_{T^*,j}$, and $\beta_{T^*,j}$ are the proportion, HIV prevalence, incidence, MDRI, and FRR of subtype $j$, respectively. This result indicates that $\Tilde{\lambda}^{\text{sub}}_{\text{old}}$ is not consistent for the average incidence in the cross-sectional population.

% {\color{red} In Appendix \ref{app: proofs}}, we derive the limiting distribution of this estimator and find that it is not consistent for the average incidence in the cross-sectional population.

%We investigated another intuitive incidence estimator that considers the number of HIV positive and negative individuals of different subtypes separately,
%\begin{align}
%\label{eq: incid estimator 2 subtype}
%    \Tilde{\lambda}^{(2)}=\frac{N_{rec}-\sum_{j=1}^K N_{\text{pos,test},\text{}j}\hat{\beta}_{T^*,j}}{\sum_{j=1}^K \frac{N_{\text{pos,test,}j}}{N_{\text{pos,}j}}N_{\text{neg}j}(\hat{\Omega}_{T^*,j}-\hat{\beta}_{T^*,j}T^*)}
%\end{align}
%, where $N_{\text{pos,test},j}$ is the number of individuals in the cross-sectional sample that are HIV positive, took the recency test and belong to the group of subtype $j$ and $N_{\text{neg},j}$ and $N_{\text{neg},j}$ are the number of HIV-negative and positive individuals for subtype $j$, respectively. Nevertheless, this estimator is also inconsistent and details on the proof are in Appendix \ref{app: proofs}.

Let $N_{\text{neg},j}$, $N_{\text{pos},j}$, and $N_{\text{rec},j}$ be the numbers of HIV-negative individuals, HIV-positive individuals, and test-recent individuals, respectively, in the cross-sectional sample that are associated with HIV subtype $j$.
We propose another estimator
\begin{align}
    \Tilde{\lambda}^{\text{sub}}=\sum_{j=1}^K \hat{\pi}_j\frac{N_{\text{rec},j} - N_{\text{pos},j}\hat{\beta}_{T^*,j}}{N_{\text{neg},j}(\hat{\Omega}_{T^*,j}-\hat{\beta}_{T^*,j}T^*)}.
    \label{eq: incidence_subtype}
\end{align}
This estimator estimates the incidences associated with each subtype separately and weights them by their sample proportions, which is consistent for the average incidence in the cross-sectional population. Details of proofs appear in Appendix \ref{app: proofs of cross-sectional subtype}.

\subsubsection{Accounting for population heterogeneity}
\label{subsubsec: incidence two population subtype}

To further incorporate population heterogeneity when both the cross-sectional and target populations have mixed subtypes, we propose the following estimators
\begin{align}
\label{eq: target background incidence estimator subtype}
    \hat{\lambda}^{\text{sub}}_{\text{ext}}=&\sum_{j=1}^K\hat{\pi}_j^*\frac{N^{-1}\sum_{i=1}^N \homega_{\text{ext},j}(\X_i) I(U_i=j)D_i(R_i-\hat{\beta}_{T^*,j})}{ N^{-1}\sum_{i=1}^N \homega_{\text{ext},j}(\X_i)I(U_i=j)(1-D_i)(\hat{\Omega}_{T^*,j}- \hat{\beta}_{T^*,j} T^*)}
\end{align}
and 
\begin{align}
\label{eq: random trial placebo incidence estimator subtype}
    \hat{\lambda}^{\text{sub}}_{\text{int}}=&\sum_{j=1}^K\hat{\pi}_j^*\frac{N^{-1}\sum_{i=1}^N \homega_{\text{int},j}(\X_i) I(U_i=j)D_i(R_i-\hat{\beta}_{T^*,j})}{ N^{-1}\sum_{i=1}^N \homega_{\text{int},j}(\X_i)I(U_i=j)(1-D_i)(\hat{\Omega}_{T^*,j}- \hat{\beta}_{T^*,j} T^*)},
\end{align}
where $U_i$ denotes the subtype that individual $i$ is associated with, $\hat{\pi}_j^* = M_j/M$, $M$ and $M_j$ are the sizes of the data from the target population and those in the data from the target population at risk for HIV subtype $j$, respectively, $\homega_{\text{ext},j}$ and $\homega_{\text{int},j}$ are estimators for the subtype-specific weights 

\begin{align}
        \omega_{\text{ext},j}(\x)=\frac{\P{}{S_i=1|\X_i=\x, U_i=j}}{\P{}{S_i=0,D_i=0|\X_i=\x, U_i=j}},
    \end{align}
and \begin{align}
        \omega_{\text{int},j}(\x)=\P{}{S_i=1|D_i=0,\X_i=\x, U_i=j}.
    \end{align}
The proofs for the consistency of $\hat{\lambda}^{\text{sub}}_{\text{ext}}$ and $\hat{\lambda}^{\text{sub}}_{\text{int}}$ can be found in Appendix \ref{app: proofs of incidence two population subtype}.

\section{Simulation study}

In this section, we will evaluate the numerical performance of the proposed estimators in realistic simulated settings, and compare it with that of standard estimators. 
In Section \ref{subsec: simulation cross-sectional samples}, we describe the general simulation strategy, including how to simulate the cross-sectional sample, estimates for MDRI and FRR to be applied in incidence estimation, and data from a target population.
When the data is of the same subtype, we evaluate the performance of the proposed extended estimators (\ref{eq: incidence_external_target}) and (\ref{eq: incidence_internal_target}) in Sections \ref{subsec: simulation external target} and \ref{subsec: simulation active-arm trial}, respectively.
We will also consider the setting where the cross-sectional population (and the target population) is a mixture of individuals at risk for two HIV subtypes. 
The proposed incidence estimators (\ref{eq: incidence_subtype}), (\ref{eq: target background incidence estimator subtype}), and (\ref{eq: random trial placebo incidence estimator subtype}) will be evaluated in Section \ref{subsec:simulation_subtype}.

\subsection{General simulation strategy}
\label{subsec: simulation cross-sectional samples}
We consider the simulation setting that mimics a men-who-have-sex-with-men (MSM) population, which is a common population of interest for HIV prevention.
Four binary covariates are considered to mimic variables associated with both HIV prevalence and incidence for MSM, including rectal infections \citep{mwaniki2023hiv, thienkrua2018young, cheung2016hiv}, receptive anal sex \citep{mmbaga2018hiv, solomon2015high, van2013evidence}, unprotected anal sex \citep{qi2015high, zhang2016hiv}, and college education \citep{solomon2015high, zhang2016hiv}.
We specify the HIV prevalence and incidence for individuals with different combinations of those variables based on the average incidences and risk ratios reported in the literature, which is given in Appendix Table \ref{tbl: general simulation strategy}.

For an individual who has been infected for $s$ units of time, the recency test result is generated from a Bernoulli distribution with success probability $\phi(s) = \text{1}- F_{\text{Gamma}}(s;\alpha=\text{11.40},\beta=\text{23.66})$, where $F_{\text{Gamma}}(\cdot;\alpha,\beta)$ is the cumulative distribution function of a Gamma random variable with shape $\alpha$ and rate $\beta$. This recency test has an MDRI of 176 days and an FRR of 0\% with $T^* = \text{2}$, and mimics a commonly applied recency test algorithm (LAg Avidity (Sedia HIV-1 LAg Avidity EIA; Sedia Biosciences Corporation, Portland, OR, USA) ODn $<\text{1.5}$ and viral load $>\text{1,000}$ copies/mL) for HIV subtype B individuals discussed in \cite{sempa2019performance}.

In each simulation, our data generation consists of the following steps:
\begin{enumerate}
    \item We simulate the cross-sectional data with size $N$ that includes individuals' covariates, HIV status, and recency test results.
    Specifically, the covariates are generated from a multivariate Bernoulli distribution with parameters given in Appendix Table \ref{tbl: general simulation strategy}.
    Individuals' HIV status at cross-sectional time, (unobserved) HIV infection duration, and recency test results for HIV-infected individuals are simulated based on the procedure discussed in Section 3.3.2 of \cite{gao2022statistical} with prevalence and incidence specified in Appendix Table \ref{tbl: general simulation strategy}.
    
    %To generate cross-sectional recency testing data with size $N$, we first simulate the covariates for each individual. 
    %Individuals' HIV status, HIV infection duration, and recency test results are then simulated, with details given in Appendix \ref{app: simulation}.
    % \item We simulate an external study (with size $m$) that includes HIV-infected individuals with known infection time and longitudinal measures of recency test results and estimate MDRI and FRR based on the study.
    % The procedure is the same as in \cite{gao2022statistical}.
    
    \item We simulate the estimates for MDRI and FRR following the procedure in Section 3.3.1 of \cite{gao2022statistical}. Specifically, we estimate MDRI based on a simulated dataset with 175 individuals and approximately 1,000 longitudinal measures of recency test results; we estimate FRR based on another simulated dataset with 1,500 individuals that are infected longer than $T^*$. 
    \item We simulate the target population of size $M$ that includes individuals' covariates only.
    The procedures are slightly different when the target population is external versus internal and will be described in the following sections.
    %We simulate a sample of individuals from the external target population or enrolled in the active-arm trial, with details described in the following sections.
    \item We apply proposed estimators to estimate the incidence in the target population, with cross-sectional data generated in (a), MDRI and FRR estimated in (b), and the target population data in (c).
    \item We estimate the variability of the estimate using two different bootstrapping strategies: a non-parametric bootstrap approach when data from the external study is available for researchers and a parametric bootstrap approach when only estimates and confidence intervals for $\beta_{T^*}$ and $\Omega_{T^*}$ are available.
    Specifically, we assume $\hat{\beta}_{T^*}$ and $\hat{\Omega}_{T^*}$ follow log-normal distribution 
    in the parametric bootstrap approach.
    Confidence intervals are estimated based on the logarithms of bootstrap estimates to account for the asymmetry of the distribution of the extended incidence estimator, with 500 bootstrap rounds. 
     
\end{enumerate}
We consider two settings: Setting 1 with $(N,M) = (\text{5,000}, \text{2,500})$ and Setting 2 with $(N,M) = (\text{2,500}, \text{1,000})$. Based on 500 simulations, we will summarize the results using the empirical mean bias, empirical standard error, mean standard error estimate, and empirical coverage probability of the 95\% confidence intervals. The ground truth that we compare our estimators with is the average incidence among individuals in the target population.

\subsection{External target population}
\label{subsec: simulation external target}
We first consider the simulation when the target population is external to the cross-sectional population, where the covariates from the target population are simulated independently with the cross-sectional population and follow the distribution given in Appendix Table \ref{tbl: general simulation strategy}.
%For a cross-sectional population with size $N$ and a target population with size $M$, we simulate the individual covariate values following the distributions given in Table \ref{tbl: general simulation strategy}. HIV status and recency test results for individuals in the cross-sectional population and estimates for MDRI and FRR are simulated as described in Section \ref{subsec: simulation cross-sectional samples}.
The true background incidence of the target population is  $\lambda=\text{0.0259}$.

We apply the proposed incidence estimator \eqref{eq: incidence_external_target} to estimate the incidence for the target population and compare it with the original incidence estimator \eqref{eq: original cross sectional incidence estimator} that does not account for population heterogeneity.
%Then based on the distribution of covariates given by HPTN 083 \citep{landovitz2021cabotegravir}, we can simulate the value of covariates of each individual in the sample from the target population of size $M=2,500$ and estimate the background HIV incidence of the target population.
%The standard error of the extended incidence estimator and the coverage probability of 95\%confidence intervals are estimated by bootstrap, with $500$ bootstrap rounds. 
Table \ref{tbl: simulation external target population} shows the simulation results.
The proposed estimator \eqref{eq: incidence_external_target}  is with less bias compared with the standard estimator \eqref{eq: original cross sectional incidence estimator}.
The bias of the proposed estimator decreases as sample size increases (comparing Setting 2 to Setting 1), while there is non-converging bias for the standard estimator.
The coverage probability using the proposed approach (especially that based on the parametric bootstrap approach) is closer to 95\%.
%For the proposed estimator, the standard error estimated by the non-parametric bootstrap is closer to the empirical standard error, while the coverage probability based on the parametric bootstrap is closer to 95\%.
\begin{table}[htbp]
\caption{\label{tbl: simulation external target population}Simulation results when the target population is external}
\centering
% \resizebox{\columnwidth}{!}{%
\renewcommand{\arraystretch}{1.75}
\begin{tabular}{cccccccc}
\hline
\multirow{2}{*}{\textbf{Setting}}& \multirow{2}{*}{\textbf{Estimator}} & \multirow{2}{*}{\textbf{Bias}$\mathbf{\times 100}$} &  \multirow{2}{*}{\textbf{SE}$\mathbf{\times 100}$}  & \multicolumn{2}{c}{\textbf{Non-parametric}}             & \multicolumn{2}{c}{\textbf{Parametric}}             \\
                           &    &                                                                             &  & \textbf{SEE}$\mathbf{\times 100}$ & \textbf{Cov}& \textbf{SEE}$\mathbf{\times 100}$ & \textbf{Cov} \\ \hline
\multirow{2}{*}{1} & Standard                         &    -0.32                                                                                 &  0.44             &   0.47         &  92.2       &       0.36        &      87.0               \\
 & Proposed                        &    0.05                                                                              &  0.55  &    0.56           &   97.0                  &        0.43       &    94.6                 \\\\
\multirow{2}{*}{2} & Standard                         &   -0.31                                                                            &  0.60  &     0.58          &   93.8                  &      0.50         &      92.2              \\
 & Proposed                        &     0.08                                                                           & 0.72  &     0.70          &   97.0                  &      0.62         &   94.8                                    \\ \hline
\end{tabular}%
\renewcommand{\arraystretch}{1}
% }
\floatfoot{\textit{Note.} For each combination of sample sizes, we show the empirical mean bias (Bias), the empirical standard error (SE), the mean standard error estimate (SEE), and the empirical coverage probability of the 95\% confidence intervals (Cov).}
\end{table}

\subsection{Internal target population}
\label{subsec: simulation active-arm trial}
We consider another simulation when the target population is internal to the cross-sectional population.
Specifically, we consider the scenario where individuals receive recency testing at the screening of an HIV prevention trial, and those HIV-negative will be invited to be enrolled in the trial.
Those HIV-negative individuals simulated in the cross-sectional population will be included in the target population with a probability that depends on covariates, which is given in Table \ref{tbl: general simulation strategy}.
Specifically, those probabilities were calculated such that the composition of the trial population is similar to the external target population in Section \ref{subsec: simulation external target}.
The true counterfactual placebo incidences are $\lambda=\text{0.0248}$ and $\lambda=\text{0.0251}$ in Settings 1 and 2, respectively.

%We first simulate cross-sectional recency testing data at the screening stage following the process described in Section \ref{subsec: simulation cross-sectional samples} with sample size $N$. Within each cross-sectional sample, every HIV-negative individual is enrolled in the trial with a probability given in Table \ref{tbl: general simulation strategy}. Specifically, those probabilities were calculated such that the composition of the trial population is realistic and similar to the external target population in Section \ref{subsec: simulation external target}. The number of individuals enrolled in the trial is fixed with $M$. Details are given in Appendix \ref{app: simulation}. 

In addition to the estimation of incidence in the target population as described in Section \ref{subsec: simulation cross-sectional samples}, we also consider the estimation of prevention efficacy when individuals in the trial receive the active prevention product and are followed for HIV infection.
For each individual with covariate $\x$ in the active-arm trial, the infection time is simulated from a Poisson process with incidence rate $\lambda(\x) (\text{1}-E)$, where $E=\text{0.5}$ is the prevention efficacy of the active product.
Let $n_{\text{trial}}$ and $y_{\text{trial}}$ be the number of HIV infections and person-time in the active-arm trial.
The HIV incidence in the trial can be estimated by $\hat\lambda_{\text{trial}} = n_{\text{trial}}/y_{\text{trial}}$ and the prevention efficacy of the active product can be estimated by $\text{1}- \hat\lambda_{\text{trial}}/\hat\lambda_{\text{int}}$ and $\text{1}- \hat\lambda_{\text{trial}}/\Tilde\lambda$, respectively.

%We consider the estimation of both the counterfactual placebo incidence of the trial population and the prevention efficacy of the investigated PrEP product. To illustrate the evaluation of the prevention efficacy, we simulate the infection time of each individual with the assumptions that the investigative prevention product decreases the incidence rate of every individual by pre-specified proportion and the incidence rate of each individual is constant over the period of follow-up. Similar to Section \ref{subsec: simulation external target}, we estimate the standard error and coverage probability of 95\%confidence intervals with parametric bootstrap and bootstrapping recency testing external studies separately.  

Table \ref{tbl: simulation internal target population} presents the simulation results. In the estimation of the incidence in the target population and the prevention efficacy, the proposed estimator shows less bias than the standard estimator. Using parametric bootstrap, the coverage probability of the 95\% confidence interval based on the proposed estimator is closer to the nominal level than the standard estimator.
\begin{table}[htbp]
\caption{\label{tbl: simulation internal target population}Simulation results when the target population is internal}
\centering
% \resizebox{\columnwidth}{!}{%
\renewcommand{\arraystretch}{1.75}
\begin{tabular}{cccccccc}
\hline
\multirow{2}{*}{\textbf{Setting}}& \multirow{2}{*}{\textbf{Estimator}} & \multirow{2}{*}{\textbf{Bias}$\mathbf{\times 100}$} &  \multirow{2}{*}{\textbf{SE}$\mathbf{\times 100}$}  & \multicolumn{2}{c}{\textbf{Non-parametric}}             & \multicolumn{2}{c}{\textbf{Parametric}}             \\
                           &    &                                                                             &  & \textbf{SEE}$\mathbf{\times 100}$ & \textbf{Cov}& \textbf{SEE}$\mathbf{\times 100}$ & \textbf{Cov} \\ \hline
\multicolumn{8}{l}{\textbf{Incidence estimation}}\\
\multirow{2}{*}{1} & Standard                     & -0.23                                                                                    & 0.48              &     0.47       &    94.0     &       0.35        &       90.4              \\
 & Proposed                       &   0.04                                                                               & 0.53   &    0.53           &   96.8                  &   0.40           &   94.4                  \\\\
\multirow{2}{*}{2} & Standard                        &  -0.23                                                                             & 0.59   &    0.59          &   96.8                  &       0.49        &    93.4                 \\
 & Proposed                        &    0.08                                                                            & 0.68  &      0.68         &   95.6                  &   0.57            &    93.6                                   \\ \hline
 \multicolumn{8}{l}{\textbf{Prevention efficacy estimation}}\\
\multirow{2}{*}{1} & Standard                       & -6.66                                                                                    & 12.29               &    13.08        &  93.8       &    11.48           &       90.6             \\
 & Proposed                        &         -0.63                                                                      &   11.06 &     11.84          &    96.6                 &    10.44          &  94.2                   \\\\
\multirow{2}{*}{2} & Standard                        & -8.31                                                                                    & 18.42               &    19.78        &  94.2       &   18.75            &     92.6                \\
 & Proposed                        &  -1.60                                                                                &   16.78 &      17.87         &    95.8                 &    17.01         &   94.6                \\ \hline
\end{tabular}%
\renewcommand{\arraystretch}{1}
% }
\end{table}

\subsection{Incorporating HIV subtypes}\label{subsec:simulation_subtype}
Finally, we consider simulation settings when there is a mixture of HIV subtypes, mimicking a mixture of subtypes A and B.
The characteristics of recency testing for individuals susceptible to subtype B have been described in Section \ref{subsec: simulation cross-sectional samples}.
For subtype A, the recency test result is generated from a Bernoulli distribution with success rate $\phi_A(s) = (\text{1}- F_{\text{Gamma}}(s;\alpha=\text{0.84},\beta=\text{1.66}))I(s\le \text{2})+\text{0.026}I(s>\text{2})$, which matches the MDRI (186 days) and FRR (2.6\%) reported in \cite{sempa2019performance}.
%In the scenario where we incorporate HIV subtype, subtypes A and B appear in the samples. Aside from subtype B, we mimic the recency test discussed in Section \ref{subsec: simulation cross-sectional samples} for subtype A, which has an MDRI and FRR of $186$ days and $2.6\%$, respectively. Specifically, for an individual infected $s$ units of time ago, the success probability for the Bernoulli distributed recency test result is $\phi(s) = (1- F_{\text{Gamma}}(s;\alpha=0.84,\beta=1.66))I(s\le 2)+0.026I(s>2)$. 

We consider the setting when the cross-sectional population includes equal numbers of individuals from subtypes A and B. In our simulation, the incidence and prevalence corresponding to subtype A are 0.02 and 0.25, respectively, and the incidence and prevalence of subtype B are 0.05 and 0.15, respectively, which leads to a true average incidence of 0.035. The simulation results applying the proposed estimator (\ref{eq: incidence_subtype}) and the standard estimator (\ref{eq: incid estimator 1 subtype}) estimating the incidence in the cross-sectional population, are given in Table \ref{tab: simulation setting incidence estimator subtype}.
The bias of the proposed estimator is smaller than the standard estimator and the coverage probability corresponding to the proposed estimator is closer to 95\% than the standard estimator.
\begin{table}[htbp]
\caption{Simulation results for incidence estimation in cross-sectional population with mixed subtypes\label{tab: simulation setting incidence estimator subtype}}
\centering
% \resizebox{\columnwidth}{!}{%
\renewcommand{\arraystretch}{1.75}
\begin{tabular}{cccccccc}
\hline
\multirow{2}{*}{\textbf{Setting}}& \multirow{2}{*}{\textbf{Estimator}} & \multirow{2}{*}{\textbf{Bias}$\mathbf{\times 100}$} &  \multirow{2}{*}{\textbf{SE}$\mathbf{\times 100}$}  & \multicolumn{2}{c}{\textbf{Non-parametric}}             & \multicolumn{2}{c}{\textbf{Parametric}}             \\
                           &    &                                                                             &  & \textbf{SEE}$\mathbf{\times 100}$ & \textbf{Cov}& \textbf{SEE}$\mathbf{\times 100}$ & \textbf{Cov} \\ \hline
\multirow{2}{*}{1} & Standard                     & 0.33                                                                                    & 0.53              &     0.56       &   90.6      &        0.51       &     88.2                \\
 & Proposed                       &   0.01                                                                               & 0.58   &    0.64           &   97.0                  &    0.52         &  94.4                   \\\\
\multirow{2}{*}{2} & Standard                        & 0.37                                                                              &  0.72  &  0.76            &   93.0                  &    0.71           &   90.4                  \\
 & Proposed                        & 0.05               &   0.75            &   0.82                  &  97.4             &  0.72 &     96.6                                \\ \hline
\end{tabular}%
\renewcommand{\arraystretch}{1}
% }
\end{table}

We also consider the estimation of incidence in the target population which also consists of multiple subtypes. The incidence, prevalence, proportions in the cross-sectional and target populations, and the enrollment probability of each combination of covariates and subtypes are given in Appendix Table \ref{tbl: general simulation strategy subtype}. The data generation and analysis follow a similar process as described in Sections \ref{subsec: simulation cross-sectional samples}, \ref{subsec: simulation external target}, and \ref{subsec: simulation active-arm trial}. The true background incidence of the target population is  $\lambda=\text{0.0269}$ when the target population is external to the cross-sectional population, while the true counterfactual placebo incidences are $\lambda=\text{0.0252}$ and $\lambda=\text{0.0264}$ in Settings 1 and 2, respectively, when the target population is internal.
% The proportions of individuals of different subtypes in the target population are assumed the same as those in the cross-sectional populations.
%simulate samples from recency testing and from the external target population or enrolled in the active-arm trial following a similar process as in Sections \ref{subsec: simulation cross-sectional samples}, \ref{subsec: simulation external target}, and \ref{subsec: simulation active-arm trial} to assess the performance of the extended incidence estimator incorporating HIV subtypes. The joint distributions of covariates and subtype in the cross-sectional population and external target or active-arm trial population can be found in Table \textcolor{red}{xxx} in the Supplementary Materials. The HIV prevalence and incidence for individuals with different categories of covariates and subtypes, are given in Table \textcolor{red}{xxx} in the Supplementary Materials. 
Table \ref{tbl: simulation two population subtype} gives the results comparing the standard estimator \eqref{eq: incidence_subtype} and the proposed estimators \eqref{eq: target background incidence estimator subtype} and \eqref{eq: random trial placebo incidence estimator subtype} in the settings where the target population is external and the trial is active-arm design, respectively. Similar to the results in Sections \ref{subsec: simulation external target} and \ref{subsec: simulation active-arm trial}, the bias of the proposed estimator is less than that of the standard estimator, and the coverage probability of the 95\% confidence interval built on the proposed estimator is either closer to 95\% or comparable with the coverage probability corresponding to the standard estimator.

\begin{table}[H]
\caption{\label{tbl: simulation two population subtype}Simulation results for two heterogeneous populations incorporating HIV subtypes}
\centering
% \resizebox{\columnwidth}{!}{%
\renewcommand{\arraystretch}{1.75}
\begin{tabular}{ccccccccc}
\hline
\multirow{2}{*}{\textbf{Setting}}& \multirow{2}{*}{\textbf{Estimator}} & \multirow{2}{*}{\textbf{Bias}$\mathbf{\times 100}$} &  \multirow{2}{*}{\textbf{SE}$\mathbf{\times 100}$}  & \multicolumn{2}{c}{\textbf{Non-parametric}}             & \multicolumn{2}{c}{\textbf{Parametric}}             \\
                          &   &                                                                          &  & \textbf{SEE}$\mathbf{\times 100}$ & \textbf{Cov}& \textbf{SEE}$\mathbf{\times 100}$ & \textbf{Cov} \\ \hline
\multicolumn{8}{l}{\textbf{External target population - incidence estimation}}\\
\multirow{2}{*}{1} & Standard                         &    -0.35                                                                                 &  0.50             &   0.48         &   93.6      &       0.40        &  87.6                   \\
& Proposed                        &    0.01                                                                              &  0.59  &    0.57          &   97.2                  &        0.48       &     94.6                \\\\
\multirow{2}{*}{2} & Standard                         &   -0.36                                                                            &  0.65  &     0.62          &  97.0                   &      0.55         &     95.0                \\
& Proposed                        &     -0.01                                                                           & 0.79  &     0.75          &  96.6                   &      0.68         &    94.8                                   \\ \hline
\multicolumn{8}{l}{\textbf{Internal target population - incidence estimation}}\\\\
\multirow{2}{*}{1} & Standard                         &   -0.13                                                                                  &  0.49             &  0.48          &  97.0       &   0.40            & 94.8                    \\
& Proposed                        &    0.05                                                                             &  0.53  & 0.51              &  97.0                   &   0.44            &  94.4                   \\\\
\multirow{2}{*}{2} & Standard                         &   -0.29                                                                            & 0.67   &  0.61             &   95.2                  &  0.56             & 95.0                    \\
& Proposed                        & 0.00                                                                               & 0.78  &   0.70            &  95.8                   &    0.64          &   95.2                                   \\ \hline
\multicolumn{8}{l}{\textbf{Internal target population - prevention efficacy estimation}}\\
\multirow{2}{*}{1} & Standard                         &   -3.82                                                                                  &   11.66            &   12.74         &      96.0   &     11.99          &      94.8               \\
& Proposed                        &   -0.31                                                                              &  11.15  &      12.02         &     96.0                &   11.37            &  94.0                   \\\\
\multirow{2}{*}{2} & Standard                         &    -8.94                                                                           & 20.74   &   19.67            &   95.6                  &   19.13            & 95.0                     \\
& Proposed                        &  -2.73                                                                              & 18.34  &    17.75           &   94.6                  &    17.24          &   94.4                                   \\ \hline
\end{tabular}%
\renewcommand{\arraystretch}{1}
% }
\end{table}
\section{Data application}
\label{sec: data application}
To further demonstrate the application and performance of the proposed methods, we analyze the data from the Sabes Study, which aims at evaluating a treatment-as-prevention intervention among cisgender men who have sex with men and transgender women in Lima, Peru \citep{lama2018design}.
In step 1 of the study, 3,337 high-risk participants were recruited and tested between July 2013 and September 2015.
Since 228 of the participants were tested for HIV in the Peruvian Biobehavioral Survey and only HIV-uninfected individuals were referred to the Sabes study, we exclude them from our analysis.
Our cross-sectional population thus includes 3,109 participants, out of which 643 were HIV-positive.
614 of the HIV-positive individuals took the recency test, which classifies an individual as recently infected if their Maxim HIV-1 LAg Avidity DBS EIA ODn < 1.5 and viral load $>\text{1,000}$ copies/mL.
This recency test has an MDRI of 202 days and an FRR of 0\% with $T^*=\text{2}$ for subtype B, which is the most prevalent HIV-1 variant in Latin America \citep{junqueira2016hiv}. 
%HIV-uninfected individuals were enrolled into the cohort after screening and re-tested monthly for incident HIV infection, with a 2-year period of follow-up. Individuals with acute or recent HIV were invited to a 48-week randomized study comparing immediate antiretroviral therapy (ART) versus deferred ART
%, where acute infection was defined as a positive HIV-RNA test in an individual with a negative HIV-antibody test and recent infection was defined as HIV diagnosed with a negative HIV-antibody or HIV-RNA test within the previous 3 months
%. 3,337 high-risk individuals were screened for HIV, 3,109 of which were identified through general clinic screening and we perform our data analysis on the individuals from general screening.

%Of the 3,109 individuals at the screening stage, 643 were HIV-positive, 614 HIV-positive individuals took recency test, 139 individuals were recency test positive, and 1,912 HIV-negative individuals were enrolled for monthly HIV testing. Here, the recency test used the Maxim HIV-1 LAg Avidity DBS EIA and classified an individual as recently infected if their LAg Avidity ODn < 1.5 and viral load > 1,000 copies/mL. This recency test has an MDRI of 202 days and an FRR of $0\%$ with $T^*=2$ for subtype B, which is the most prevalent HIV-1 variant in Latin America \citep{junqueira2016hiv}.

In step 2 of the study, 1,912 HIV-negative individuals were enrolled for monthly HIV testing with a 2-year period of follow-up. 
Our goal is to estimate the HIV incidence for the step 2 study population using the recency testing data collected in step 1 of the study.
We consider risk factors that potentially modify HIV prevalence and incidence, including age (18-22, 23-26, 27-32, or >32), sexuality (homosexual, bisexual, heterosexual, or transgender), sexual role (insertive, receptive, or versatile), postsecondary education (yes or no), number of sex partners (0, 1, 2-5, or >5), anal sex without condom use (yes or no), anal or vaginal sex with HIV positive partners (yes, no, or do not know), whether an individual has a main partner (yes or no), transactional sex (yes or no), and alcohol use (yes or no). The factors related to sexual behavior and alcohol use were measured based on individuals' behaviors in the 3 months preceding the screening.
To account for the fact that not all HIV-positive individuals took the recency test, we apply modified versions of \eqref{eq: original cross sectional incidence estimator} and \eqref{eq: incidence_internal_target}, which are given by
\begin{align}
\label{eq: modified original cross sectional incidence estimator in body text}
    \Tilde{\lambda}^{\text{mod}}=&\frac{N_{\text{rec}}-N_{\text{pos,test}}\hat{\beta}_{T^*}}{N_{\text{pos}}^{-1}N_{\text{pos,test}} N_{\text{neg}}(\hat{\Omega}_{T^*}- \hat{\beta}_{T^*} T^*)},
\end{align}
and
\begin{align} 
\label{eq: modified incidence_internal_target in body text}
\hat{\lambda}_{\text{int}}^{\text{mod}}=&\frac{N^{-1}\sum_{i=1}^N \homega_{\text{int}}(\X_i) D_iQ_i(R_i-\hat{\beta}_{T^*})}{N_{\text{pos}}^{-1} N_{\text{pos,test}}N^{-1}\sum_{i=1}^N \homega_{\text{int}}(\X_i)(1-D_i)(\hat{\Omega}_{T^*}- \hat{\beta}_{T^*} T^*)},
\end{align}
respectively, where $Q_i$ is the indicator of receiving recency test for an HIV-positive individual $i$ and $N_{\text{pos,test}} =\sum_{i=1}^ND_iQ_i$ is the number of HIV-positive individuals who took the recency test. 
% Details on the modification are given in Appendix \ref{app: modified estimators}.
We assume that the recency testing is missing at random, which holds in our analysis. Indeed, the recency testing is missing completely at random in our setting since the missing recency testing is due to insufficient blood samples and/or inadequate storage.
Similar to the simulation studies, we use both parametric and non-parametric bootstrap to estimate the confidence interval with 500 bootstrap rounds.

% In the cross-sectional population, 139 individuals were recency test positive, such that the estimated incidence in the step 1 population is {\color{red}xxx (xxx,xxx)}.
% The incidences of the cohort population estimated by \eqref{eq: modified original cross sectional incidence estimator in body text} and \eqref{eq: modified incidence_internal_target in body text} are 0.087 (parametric 95\% CI: (0.074,0.103); non-parametric 95\% CI: (0.064,0.119)) and 0.089 per person-year (parametric 95\% CI: (0.075,0.105); non-parametric 95\% CI: (0.065,0.121)), respectively. 
% The incidences estimated by \eqref{eq: modified original cross sectional incidence estimator in body text} and \eqref{eq: modified incidence_internal_target in body text} are very similar because a large proportion (78\%) of HIV-negative individuals were enrolled to the target population and the distributions of risk factors in the screening and target populations are very similar. This can also be seen through a narrow range of the estimated weight $\homega_{\text{int}}$ (0.53, 0.93).
The data application result is now under review by the Sabes study team. We will update the result once it gets approved.
% Using the data for the infection time of individuals in the monthly tested cohort, the estimated incidence of the cohort population is 0.11 per person-year, which is comparable with the estimation by \eqref{eq: modified original cross sectional incidence estimator in body text} and \eqref{eq: modified incidence_internal_target in body text}. This may be due to the fact that the only intervention for the cohort was monthly testing and thus, the background incidence was similar to the incidence during follow-up.

% , 2,109 of which were followed up for monthly HIV testing. Eventually, 216 individuals with acute or recent infection were enrolled in the randomized ART trial, including 12 individuals with incident HIV referred from a local STI clinic.

\section{Discussion}
In this paper, we outlined a statistical framework to address population heterogeneity for HIV incidence estimation based on recency test. We considered two specific settings including the extension of a cross-sectional incidence estimate to an external target population and an internal population. We derived the closed forms of extended estimators for the two settings respectively based on re-weighting \cite{kassanjee2012new} estimator. We also conducted simulation studies to extensively analyze the performance of proposed estimators in realistic simulation scenarios.

The extended estimators work well when all assumptions hold, including the two key assumptions in \cite{gao2022statistical} and the assumption that $\X$ captures all the factors that modify HIV prevalence and incidence. Possible results induced by the violation of the two key assumptions have been discussed in \cite{gao2022statistical}. Further studies should consider performing sensitivity analysis to evaluate the biases of the estimators when the assumption of no unmeasured confounding fails to hold.

% further studies can resolve is the incorporation of time-dependent covariates. Only the covariate values of HIV-negative individuals can modify HIV incidence since only HIV-negative individuals can contribute to incident HIV infections. For HIV-positive individuals, the values of their covariates may be correlated with but very different from the values when they were HIV-negative. Thus, their covariate values do not directly modify HIV incidence and adjustments should be made.

In the article, we accounted for heterogeneity from time-independent covariates in the cross-sectional and target populations. Indeed, there may be time-dependent variables, e.g., those measuring sexual behaviors, that also modify HIV prevalence and incidence. However, it may be very difficult to incorporate time-dependent covariates in the proposed framework, given that no longitudinal trajectory of the time-dependent covariates can be obtained from individuals in the cross-sectional population.
%Indeed, some of the time-independent covariates we included, e.g., the binary variable of whether there is unprotected receptive anal sex, is a proxy of the time-dependent xxx we would like to incorporate.

% \textcolor{red}{how to end our article? Like, our methods can also be applied to other areas like Covid-19 trials?}

\bibliographystyle{apalike} 
\bibliography{reference}
%-------------------------------------------------------------------------------------------------------------------------------------------------------
%Appendices

% \clearpage

\appendix
\begin{appendices}
% \titleformat{\section}{\Large\bfseries}{\thesection.}{0.5em}{}
% \titleformat{\subsection}{\large\bfseries}{\thesubsection.}{0.5em}{}
% \titleformat{\subsubsection}{\large\bfseries}{\thesubsubsection.}{0.5em}{}

% \setcounter{page}{1}

% \begin{appendices}

% \setcounter{footnote}{0}

% % \begin{center}
% % 	\textbf{\huge Online Appendices}
% % \end{center}

% %-------------------------------------------------------------------------------------------------------------------------------------------------------

% \setcounter{table}{0}
% \setcounter{figure}{0}

\section{Additional details of proofs}
\label{app: proofs}

\subsection{Proofs of Section \ref{subsubsec: cross-sectional incidence estimator subtype}}
\label{app: proofs of cross-sectional subtype}
Let $U_i$ denote the HIV subtype of individual $i$. The average incidence in the cross-sectional population is $\sum_{j=1}^K \lambda(j)\pi_j$. However, the asymptotic behavior of \eqref{eq: incid estimator 1 subtype} is given by
\begin{align*}
    \frac{N_{\text{rec}}}{N}=\frac{\sum_{i=1}^N D_iR_i}{N}\overset{\text{p}}{\rightarrow}&\, \E{}{D_iR_i}=\sum_{j=1}^K\E{}{D_iR_i|U_i=j}\P{}{U_i=j}\\
    =&\sum_{j=1}^K \sbr{(\Omega_{T^*,j}-\beta_{T^*,j}T^*)(1-p(j))\lambda(j)+p(j)\beta_{T^*,j}}\pi_j; \\
    \frac{N_{\text{pos}}}{N}=\frac{\sum_{i=1}^N D_i}{N}\overset{\text{p}}{\rightarrow}&\,\E{}{D_i}= \sum_{j=1}^K\E{}{D_i|U_i=j}\P{}{U_i=j}=\sum_{j=1}^K p(j)\pi_j;\\
    \frac{N_{\text{neg}}}{N}=\frac{\sum_{i=1}^N (1-D_i)}{N}\overset{\text{p}}{\rightarrow}&\,\E{}{(1-D_i)}
    % =\sum_{j=1}^K\E{}{(1-D_i)|U_i=j}\P{}{U_i=j}
    = 1-\sum_{j=1}^K p(j)\pi_j;\\
    \Tilde{\lambda}^{\text{sub}}_{\text{old}} \overset{\text{p}}{\rightarrow}&\, \frac{\sum_{j=1}^K \sbr{(\Omega_{T^*,j}-\beta_{T^*,j}T^*)(1-p(j))\pi_j\lambda(j)+p(j)\pi_j\beta_{T^*,j}} - \p{\sum_{j=1}^K p(j)\pi_j}\p{\sum_{j=1}^K \pi_j\beta_{T^*,j}}}{\p{1-\sum_{j=1}^K p(j)\pi_j}\p{\sum_{j=1}^K \pi_j(\Omega_{T^*,j}-\beta_{T^*,j}T^*)}},
\end{align*}
which indicates that \eqref{eq: incid estimator 1 subtype} is inconsistent for the average incidence in the cross-sectional population.

Moreover, \eqref{eq: incidence_subtype} can be reformulated as
\begin{align*}
    \Tilde{\lambda}^{\text{sub}}=\sum_{j=1}^K \hat{\pi}_j\frac{N^{-1}\sum_{i=1}^N D_i(R_i-\hat{\beta}_{T^*,j})I(U_i=j)}{N^{-1}\sum_{i=1}^N (1-D_i)(\hat{\Omega}_{T^*,j}- \hat{\beta}_{T^*,j} T^*)I(U_i=j)}:=\sum_{j=1}^K \hat{\pi}_j A_j.
\end{align*}
For the numerator of $A_j$, we have
\begin{align*}
    \frac{1}{N}\sum_{i=1}^N D_i(R_i-\hat{\beta}_{T^*,j})I(U_i=j)\overset{\text{p}}{\rightarrow}& \E{}{D_i(R_i-\hat{\beta}_{T^*,j})I(U_i=j)}\\
    =& \E{}{D_iR_iI(U_i=j)} -\E{}{D_i\hat{\beta}_{T^*,j}I(U_i=j)}\\
    =& \E{}{D_iR_i|U_i=j}\pi_j -\E{}{D_i|U_i=j}\pi_j\beta_{T^*,j}\\
    =& \sbr{(\Omega_{T^*,j}-\beta_{T^*,j}T^*)(1-p(j))\lambda(j)+p(j)\beta_{T^*,j}}\pi_j - p(j)\pi_j\beta_{T^*,j}\\
    =&(\Omega_{T^*,j}-\beta_{T^*,j}T^*)(1-p(j))\lambda(j)\pi_j.
\end{align*}
Similarly, for the denominator of $A_j$,
\begin{align*}
    \frac{1}{N} \sum_{i=1}^N (1-D_i)(\hat{\Omega}_{T^*,j}- \hat{\beta}_{T^*,j} T^*)I(U_i=j) \overset{\text{p}}{\rightarrow}& (\Omega_{T^*,j}-\beta_{T^*,j}T^*)(1-p(j))\pi_j.
\end{align*}
Therefore, we get
\begin{align*}
    \Tilde{\lambda}^{\text{sub}}=\sum_{j=1}^K \hat{\pi}_j A_j \overset{\text{p}}{\rightarrow} \sum_{j=1}^K \pi_j \lambda(j),
\end{align*}
which implies that \eqref{eq: incidence_subtype} is consistent for the average incidence in the cross-sectional population.

\subsection{Proofs of Section \ref{subsubsec: incidence two population subtype}}
\label{app: proofs of incidence two population subtype}
Let $p(\x,j)$ and $\lambda(\x,j)$ be the prevalence and incidence of an individual with covariates $\X=\x$ and subtype $j$, respectively.
Let $g(\x,j)$ and $h(\x,j)$ be the joint probability density function of covariates and subtype in the cross-sectional population and target population, respectively. Consider the estimator
\begin{align*}
    \hat{\lambda}^{\text{sub}}=&\sum_{j=1}^K\hat{\pi}_j^*\frac{N^{-1}\sum_{i=1}^N \omega_{j}(\X_i) I(U_i=j)D_i(R_i-\hat{\beta}_{T^*,j})}{ N^{-1}\sum_{i=1}^N \omega_{j}(\X_i)I(U_i=j)(1-D_i)(\hat{\Omega}_{T^*,j}- \hat{\beta}_{T^*,j} T^*)}:=\sum_{j=1}^K\hat{\pi}_j^* B_j,
\end{align*}
where $\omega_j$ is the weight corresponding to subtype $j$.
The numerator of $B_j$ converges in probability to
\begin{align*}
    &\E{}{\omega_j(\X_i)I(U_i=j)D_i(R_i-\hat{\beta}_{T^*,j})}\\
    =&\E{}{\E{}{\omega_j(\X_i)I(U_i=j)D_i(R_i-\hat{\beta}_{T^*,j})|\X_i}}\\
    =&\E{}{\omega_j(\X_i)\E{}{I(U_i=j)D_i(R_i-\hat{\beta}_{T^*,j})|\X_i}}\\
    =&\E{}{\omega_j(\X_i)\p{\E{}{I(U_i=j)D_iR_i|\X_i} - \E{}{I(U_i=j)D_i\hat{\beta}_{T^*,j}|\X_i}}}\\
    =&\E{}{\omega_j(\X_i)\p{\P{}{U_i=j,D_i=1,R_i=1|\X_i} - \beta_{T^*,j}\P{}{U_i=j,D_i=1|\X_i}}}\\
    =&\E{}{\omega_j(\X_i)(\P{}{D_i=1,R_i=1|U_i=j,\X_i} - \beta_{T^*,j}\P{}{D_i=1|U_i=j,\X_i})\P{}{U_i=j|\X_i}}\\
    =&\E{}{\omega_j(\X_i)\sbr{(\Omega_{T^*,j}-\beta_{T^*,j}T^*)\lambda(\X_i,j)(1-p(\X_i,j))+\beta_{T^*,j}p(\X_i,j) - \beta_{T^*,j}p(\X_i,j)}\P{}{U_i=j|\X_i}}\\
    =&(\Omega_{T^*,j}-\beta_{T^*,j}T^*)\E{}{\omega_j(\X_i)\lambda(\X_i,j)(1-p(\X_i,j))\P{}{U_i=j|\X_i}}.
\end{align*}
Similarly, the denominator of $B_j$ converges in probability to
\begin{align*}
    (\Omega_{T^*,j}-\beta_{T^*,j}T^*)\E{}{\omega_j(\X_i)\p{1-p(\X_i,j)}\P{}{U_i=j|\X_i}}.
\end{align*}
Furthermore, $\hat{\pi}_j^*$ converges in probability to the true proportion of subtype $j$ in the target population. This proportion is equal to $\int h(\x,j) d\x$. Therefore, if we adopt a weight of the form $\omega_j(\x)\propto \frac{h(\x,j)}{\p{1-p(\x,j)}g(\x,j)}$, we have
\begin{align*}
    \hat{\lambda}^{\text{sub}} \overset{\text{p}}{\rightarrow}& \sum_{j=1}^K \frac{\E{}{\omega_j(\X_i)\lambda(\X_i,j)(1-p(\X_i,j))\P{}{U_i=j|\X_i}}}{\E{}{\omega_j(\X_i)\p{1-p(\X_i,j)}\P{}{U_i=j|\X_i}}}\int h(\x,j) d\x\\
    =&\sum_{j=1}^K\frac{\int \omega_j(\x)\lambda(\x,j)(1-p(\x,j))\P{}{U_i=j|\x}g(\x) d\x}{\int \omega_j(\x)(1-p(\x,j))\P{}{U_i=j|\x}g(\x) d\x}\int h(\x,j) d\x\\
    =&\sum_{j=1}^K\frac{\int \omega_j(\x)\lambda(\x,j)(1-p(\x,j))g(\x,j) d\x}{\int \omega_j(\x)(1-p(\x,j))g(\x,j) d\x}\int h(\x,j) d\x\\
    =&\sum_{j=1}^K\frac{\int \frac{h(\x,j)}{\p{1-p(\x,j)}g(\x,j)}\lambda(\x,j)(1-p(\x,j))g(\x,j) d\x}{\int \frac{h(\x,j)}{\p{1-p(\x,j)}g(\x,j)}(1-p(\x,j))g(\x,j) d\x}\int h(\x,j) d\x\\
    =&\sum_{j=1}^K\frac{\int \lambda(\x,j)h(\x,j) d\x}{\int h(\x,j) d\x}\int h(\x,j) d\x\\
    =&\sum_{j=1}^K\int \lambda(\x,j)h(\x,j) d\x.
\end{align*}
This implies that $\hat{\lambda}^{\text{sub}}$ is consistent for the average incidence in the cross-sectional population. Regarding the weight estimation, when the target population is external,
\begin{align*}
    \omega_j(\x)\propto \frac{h(\x,j)}{\p{1-p(\x,j)}g(\x,j)}=&\frac{f(\x,j|S_i=1)}{\P{}{D_i=0|S_i=0,\X_i=\x, U_i=j}f(\x,j|S_i=0)}\\
    =&\frac{\P{}{S_i=1|\X_i=\x, U_i=j}f(\x,j)/\P{}{S_i=1}}{\P{}{D_i=0|S_i=0,\X_i=\x, U_i=j}\P{}{S_i=0|\X_i=\x, U_i=j}f(\x,j)/\P{}{S_i=0}}\\
    =&\frac{\P{}{S_i=1|\X_i=\x, U_i=j}}{\P{}{D_i=0|S_i=0,\X_i=\x, U_i=j}\P{}{S_i=0|\X_i=\x, U_i=j}}\frac{\P{}{S_i=0}}{\P{}{S_i=1}}\\
    \propto&\frac{\P{}{S_i=1|\X_i=\x, U_i=j}}{\P{}{S_i=0,D_i=0|\X_i=\x, U_i=j}},
\end{align*}
where we let $f(\x,j|\cdot)$ denote the generic notation for the joint probability density function of covariates and subtype in a population.
That is, we can apply the weight
\begin{align*}
% \label{eq: weight when target population is external}
    \omega_{\text{ext},j}(\x)=\frac{\P{}{S_i=1|\X_i=\x, U_i=j}}{\P{}{S_i=0,D_i=0|\X_i=\x, U_i=j}},
\end{align*}
Similarly, when the target population is internal,
\begin{align*}
    \omega_j(\x)\propto& \frac{h(\x,j)}{(1-p(\x,j))g(\x,j)}\\
    =&\frac{g(\x,j|D_i=0,S_i=1)}{\P{}{D_i=0|\X_i=\x, U_i=j}g(\x,j)}\\
    =&\frac{\P{}{D_i=0,S_i=1|\X_i=\x, U_i=j}g(\x,j)/\P{}{D_i=0,S_i=1}}{\P{}{D_i=0|\X_i=\x, U_i=j}g(\x,j)}\\
    =&\frac{\P{}{D_i=0,S_i=1|\X_i=\x, U_i=j}/\P{}{D_i=0,S_i=1}}{\P{}{D_i=0|\X_i=\x, U_i=j}}\\
    =&\frac{\P{}{S_i=1|D_i=0,\X_i=\x, U_i=j}}{\P{}{D_i=0,S_i=1}}\\
    \propto& \P{}{S_i=1|D_i=0,\X_i=\x, U_i=j}.
\end{align*}
Thus, we can adopt $\omega_{\text{int},j}(\x)=\P{}{S_i=1|D_i=0,\X_i=\x, U_i=j}$ as the weight.

\clearpage
\section{Additional details of simulation study}
\label{app: simulation}

\subsection{Parameter specification for simulations}
\label{app: parameter spec simulation}
\begin{table}[hbt!]
\caption{\label{tbl: general simulation strategy}Parameter specification for simulations with single HIV subtype}
\centering
\resizebox{\columnwidth}{!}{%
\renewcommand{\arraystretch}{1.75}
\begin{tabular}{cccccccccc}
\hline
%\multirow{2}{*}{Rectal infection} & \multirow{2}{*}{Receptive} & \multirow{2}{*}{Unprotected} & \multirow{2}{*}{College} & \multirow{2}{*}{Incidence} & \multirow{2}{*}{Prevalence} & \multirow{2}{*}{Cross-sectional proportion} & \multirow{2}{*}{Target proportion} & \multicolumn{2}{c}{Enrollment probability} \\
%                                  &                            &                              &                          &                            &                             &                                                                                       &                                                                              & Setting 1               & Setting 2              \\ \hline
\textbf{Rectal} & \textbf{Receptive} & \textbf{Unprotected} & \multirow{2}{*}{\textbf{College}} & \multirow{2}{*}{\textbf{Incidence}} & \multirow{2}{*}{\textbf{Prevalence}} & \multicolumn{2}{c}{\textbf{Proportion}}  & \multicolumn{2}{c}{\textbf{Enrollment probability}} \\
\textbf{infection} &          \textbf{anal sex}                  &                \textbf{anal sex}              &                          &                            &                             &      \textbf{Cross-sectional pop.}                                                                                 &       \textbf{Target pop.}                                                                       & \textbf{Setting 1}               & \textbf{Setting 2}              \\ \hline
0                                 & 0                          & 0                            & 0                        & 0.005                      & 0.092                       & 0.033                                                                                 & 0.016                                                                        & 0.256               & 0.205                \\
0                                 & 0                          & 0                            & 1                        & 0.003                      & 0.057                       & 0.094                                                                                 & 0.064                                                                        & 0.364               & 0.291                \\
0                                 & 0                          & 1                            & 0                        & 0.024                      & 0.135                       & 0.016                                                                                 & 0.040                                                                        & 0.941               & 0.849                \\
0                                 & 0                          & 1                            & 1                        & 0.012                      & 0.083                       & 0.104                                                                                 & 0.111                                                                        & 0.584               & 0.467                \\
0                                 & 1                          & 0                            & 0                        & 0.009                      & 0.148                       & 0.033                                                                                 & 0.029                                                                        & 0.527               & 0.421                \\
0                                 & 1                          & 0                            & 1                        & 0.005                      & 0.092                       & 0.097                                                                                 & 0.118                                                                        & 0.667               & 0.533                \\
0                                 & 1                          & 1                            & 0                        & 0.043                      & 0.217                       & 0.133                                                                                 & 0.102                                                                        & 0.488               & 0.390                \\
0                                 & 1                          & 1                            & 1                        & 0.022                      & 0.134                       & 0.359                                                                                 & 0.341                                                                        & 0.548               & 0.438                \\
1                                 & 0                          & 0                            & 0                        & 0.013                      & 0.176                       & 0.003                                                                                 & 0.001                                                                        & 0.243               & 0.194                \\
1                                 & 0                          & 0                            & 1                        & 0.007                      & 0.109                       & 0.008                                                                                 & 0.005                                                                        & 0.315               & 0.252                \\
1                                 & 0                          & 1                            & 0                        & 0.063                      & 0.259                       & 0.001                                                                                 & 0.003                                                                        & 0.950               & 0.950                \\
1                                 & 0                          & 1                            & 1                        & 0.032                      & 0.160                       & 0.006                                                                                 & 0.012                                                                        & 0.932               & 0.914                \\
1                                 & 1                          & 0                            & 0                        & 0.024                      & 0.284                       & 0.004                                                                                 & 0.008                                                                        & 0.944               & 0.879                \\
1                                 & 1                          & 0                            & 1                        & 0.012                      & 0.176                       & 0.016                                                                                 & 0.025                                                                        & 0.944               & 0.755                \\
1                                 & 1                          & 1                            & 0                        & 0.114                      & 0.417                       & 0.028                                                                                 & 0.042                                                                        & 0.937               & 0.805                \\
1                                 & 1                          & 1                            & 1                        & 0.057                      & 0.258                       & 0.064                                                                                 & 0.084                                                                        & 0.882               & 0.706                \\ \hline
\end{tabular}%
\renewcommand{\arraystretch}{1}
}
\floatfoot{\textit{Note.} This table presents the incidence, prevalence, proportions in the cross-sectional and target populations, and enrollment probability of each category of individuals. Enrollment probabilities were calculated such that the distribution of covariates in the internal target population is similar to the distribution given by the proportions of covariates in the target population shown in this table. The enrollment probabilities in Setting 1 are different from those in Setting 2 since the sample sizes are different in the two settings. Details for generating the enrollment probabilities are given in Appendix \ref{app: simulation enrollment probability}.}
\end{table}

\begin{table}[hbt!]
\caption{\label{tbl: general simulation strategy subtype}Parameter specification for simulations with multiple HIV subtypes}
\centering
\resizebox{\columnwidth}{!}{%
\renewcommand{\arraystretch}{1.75}
\begin{tabular}{ccccccccccc}
\hline
\textbf{Rectal} & \textbf{Receptive} & \textbf{Unprotected} & \multirow{2}{*}{\textbf{College}} & \multirow{2}{*}{\textbf{Subtype}} & \multirow{2}{*}{\textbf{Incidence}} & \multirow{2}{*}{\textbf{Prevalence}} & \multicolumn{2}{c}{\textbf{Proportion}}  & \multicolumn{2}{c}{\textbf{Enrollment probability}} \\
\textbf{infection} &          \textbf{anal sex}                  &                \textbf{anal sex}              &    &                      &                            &                             &      \textbf{Cross-sectional pop.}                                                                                 &       \textbf{Target pop.}                                                                       & \textbf{Setting 1}               & \textbf{Setting 2}              \\ \hline
0              & 0         & 0             & 0       & B       & 0.0056    & 0.0689     & 0.0163     & 0.0110     & 0.3614              & 0.2891               \\
0              & 0         & 0             & 0       & A       & 0.0045    & 0.1148     & 0.0171     & 0.0046     & 0.1505              & 0.1204               \\
0              & 0         & 0             & 1       & B       & 0.0028    & 0.0426     & 0.0441     & 0.0466     & 0.5512              & 0.4409               \\
0              & 0         & 0             & 1       & A       & 0.0023    & 0.0710     & 0.0497     & 0.0178     & 0.1929              & 0.1543               \\
0              & 0         & 1             & 0       & B       & 0.0266    & 0.1011     & 0.0099     & 0.0256     & 0.8872              & 0.9434               \\
0              & 0         & 1             & 0       & A       & 0.0213    & 0.1685     & 0.0064     & 0.0141     & 0.8843              & 0.9431               \\
0              & 0         & 1             & 1       & B       & 0.0135    & 0.0626     & 0.0527     & 0.0771     & 0.7810              & 0.6248               \\
0              & 0         & 1             & 1       & A       & 0.0107    & 0.1043     & 0.0510     & 0.0338     & 0.3700              & 0.2960               \\
0              & 1         & 0             & 0       & B       & 0.0101    & 0.1111     & 0.0154     & 0.0215     & 0.7826              & 0.6261               \\
0              & 1         & 0             & 0       & A       & 0.0081    & 0.1851     & 0.0171     & 0.0078     & 0.2779              & 0.2223               \\
0              & 1         & 0             & 1       & B       & 0.0051    & 0.0688     & 0.0527     & 0.0812     & 0.8281              & 0.6625               \\
0              & 1         & 0             & 1       & A       & 0.0041    & 0.1146     & 0.0445     & 0.0365     & 0.4629              & 0.3704               \\
0              & 1         & 1             & 0       & B       & 0.0479    & 0.1631     & 0.0651     & 0.0653     & 0.5990              & 0.4792               \\
0              & 1         & 1             & 0       & A       & 0.0383    & 0.2718     & 0.0681     & 0.0365     & 0.3682              & 0.2946               \\
0              & 1         & 1             & 1       & B       & 0.0242    & 0.1010     & 0.1769     & 0.2483     & 0.7807              & 0.6246               \\
0              & 1         & 1             & 1       & A       & 0.0193    & 0.1683     & 0.1824     & 0.0927     & 0.3053              & 0.2442               \\
1              & 0         & 0             & 0       & B       & 0.0149    & 0.1324     & 0.0009     & 0.0009     & 0.6142              & 0.4913               \\
1              & 0         & 0             & 0       & A       & 0.0119    & 0.2206     & 0.0026     & 0.0005     & 0.1140              & 0.0912               \\
1              & 0         & 0             & 1       & B       & 0.0075    & 0.0819     & 0.0043     & 0.0037     & 0.4643              & 0.3715               \\
1              & 0         & 0             & 1       & A       & 0.0060    & 0.1366     & 0.0039     & 0.0009     & 0.1371              & 0.1097               \\
1              & 0         & 1             & 0       & B       & 0.0702    & 0.1944     & 0.0004     & 0.0027     & 0.9500              & 0.9500               \\
1              & 0         & 1             & 0       & A       & 0.0561    & 0.3240     & 0.0009     & 0.0005     & 0.3941              & 0.3153               \\
1              & 0         & 1             & 1       & B       & 0.0355    & 0.1203     & 0.0030     & 0.0096     & 0.8965              & 0.9444               \\
1              & 0         & 1             & 1       & A       & 0.0283    & 0.2005     & 0.0034     & 0.0027     & 0.4999              & 0.3999               \\
1              & 1         & 0             & 0       & B       & 0.0267    & 0.2135     & 0.0013     & 0.0055     & 0.9187              & 0.9467               \\
1              & 1         & 0             & 0       & A       & 0.0214    & 0.3559     & 0.0026     & 0.0027     & 0.8272              & 0.6618               \\
1              & 1         & 0             & 1       & B       & 0.0135    & 0.1322     & 0.0094     & 0.0183     & 0.8791              & 0.8931               \\
1              & 1         & 0             & 1       & A       & 0.0108    & 0.2203     & 0.0064     & 0.0064     & 0.6378              & 0.5103               \\
1              & 1         & 1             & 0       & B       & 0.1264    & 0.3135     & 0.0167     & 0.0246     & 0.8781              & 0.8598               \\
1              & 1         & 1             & 0       & A       & 0.1009    & 0.5225     & 0.0111     & 0.0173     & 0.8919              & 0.9439               \\
1              & 1         & 1             & 1       & B       & 0.0638    & 0.1941     & 0.0308     & 0.0584     & 0.8806              & 0.9403               \\
1              & 1         & 1             & 1       & A       & 0.0510    & 0.3234     & 0.0330     & 0.0251     & 0.5626              & 0.4501     \\\hline         
\end{tabular}%
\renewcommand{\arraystretch}{1}
}
\end{table}

\clearpage
\subsection{Generation of enrollment probability}
\label{app: simulation enrollment probability}
Suppose that we have $C$ different combinations of covariates and subtype. Let $e_c$ be the enrollment probability of the combination $c$ ($c=1,\ldots,C$). We define the raw enrollment probability $e_c^*$ as the probability that makes the joint distribution of covariates and subtype in the internal target population exactly the same as the joint distribution in the external target population. Let $g_c$, $h_c$, $p_c$ be the proportion of the combination $c$ in the cross-sectional population, the proportion of $c$ in the external target population, and the prevalence of $c$ in the cross-sectional population, respectively. Then, the raw enrollment probability is given by
\begin{align*}
    e_c^* = \frac{h_cM}{(1-p_c)g_cN}.
\end{align*}
One observation is that different sample sizes $N$ and $M$ may lead to different raw enrollment probabilities. Numerically $e_c^*$ may be greater than 1. To overcome this, we apply truncation to the raw enrollment probability to generate the final enrollment probability. Specifically, let $e^*_{m}$, $e^*_{ma}$, and $e^*_{mi}$ be the maximum raw enrollment probability that is less than $0.9$, the maximum raw enrollment probability, and the minimum raw enrollment probability that is greater than 1, respectively, and then, the enrollment probability of the combination $c$ is given by
\begin{align*}
    e_c = e^*_cI(e^*_c\le 1) + \sbr{e^*_{m}+0.05+\frac{(0.9 - e^*_{m})(e^*_c-e^*_{mi})}{e^*_{ma}-e^*_{mi}}}I(e^*_c> 1).
\end{align*}
This truncation guarantees that all the raw enrollment probabilities that are greater than 1 are truncated to no greater than 0.95 and that the order of those probabilities is kept.

\end{appendices}

\end{document}